\documentclass[11pt,cqg]{iopart}
\pdfoutput=1

\usepackage{amsopn,amsthm}
\usepackage{iopams}
\usepackage{mathrsfs}
\usepackage[numbers,square,sort&compress]{natbib}
\usepackage{graphicx}
\usepackage[pdftex]{hyperref}

\DeclareRobustCommand\hbar{\mathchar'26\mkern-9mu h}
\usepackage{pstricks}
\usepackage{microtype}
\usepackage{caption}
\usepackage{tikz}
\usetikzlibrary{shapes.geometric}
\usetikzlibrary{decorations.pathmorphing}
\usetikzlibrary{arrows,shapes,positioning}
\usetikzlibrary{decorations.markings}
\tikzstyle arrowstyle=[scale=1]
\tikzstyle directed=[postaction={decorate,decoration={markings,mark=at position .65 with {\arrow[arrowstyle]{stealth}}}}]
\tikzstyle reverse directed=[postaction={decorate,decoration={markings,mark=at position .65 with {\arrowreversed[arrowstyle]{stealth};}}}]

\def\ben{\begin{equation}}
\def\een{\end{equation}}
\def\bena{\begin{eqnarray}}
\def\eena{\end{eqnarray}}

\newcommand{\eH}{\mathscr{H}}
\newcommand{\eB}{\mathscr{B}}
\newcommand{\eC}{\mathscr{C}}
\newcommand{\M}{\mathscr{M}}
\newcommand{\D}{\mbox{d}}
\newcommand{\E}{\mathcal{E}}

\newcommand{\RR}{\mathbb{R}}
\newcommand{\CC}{\mathbb{C}}

\newcommand{\half}{\frac{1}{2}}

\renewcommand{\hbar}{h}

\newcommand{\so}{\frak{so}}
\newcommand{\h}{\frak{h}}
\newcommand{\g}{\frak{g}}
\newcommand{\ad}{{\rm ad}}

\newcommand{\bX}{{\boldsymbol{X}}}

\DeclareMathOperator{\scri}{\mathscr{I}}

\newtheorem{thm}{Theorem}
\newtheorem{lemma}{Lemma}[section]

\begin{document}

\title{Superradiant instabilities of asymptotically anti--de Sitter black holes}

\author{Stephen R. Green$^1$, Stefan Hollands$^2$, Akihiro
  Ishibashi$^3$ and Robert M. Wald$^4$}
\address{$^1$ Perimeter Institute for Theoretical Physics\\
 31 Caroline Street North, Waterloo, Ontario N2L 2Y5, Canada}
\address{$^2$ Institut f\"ur Theoretische Physik, Universit\"at Leipzig\\
 Br\"uderstrasse 16, D-04103 Leipzig, Germany}
\address{$^3$ Department of Physics, Kinki University,
 Higashi-Osaka, 577-8502, Japan}
\address{$^4$ Enrico Fermi Institute and Department of Physics, The University of Chicago\\
 5640 South Ellis Avenue, Chicago, Illinois 60637, USA}
\eads{\mailto{sgreen@perimeterinstitute.ca},
 \mailto{stefan.hollands@uni-leipzig.de},
 \mailto{akihiro@phys.kindai.ac.jp} and \mailto{rmwa@uchicago.edu}}

\begin{abstract}
  We study the linear stability of asymptotically anti--de Sitter
  black holes in general relativity in spacetime dimension
  $d\ge4$. Our approach is an adaptation of the general framework of
  Hollands and Wald, which gives a stability criterion in terms of the
  sign of the canonical energy, $\mathcal{E}$. The general framework
  was originally formulated for static or stationary and axisymmetric
  black holes in the asymptotically flat case, and the stability
  analysis for that case applies only to axisymmetric
  perturbations. However, in the asymptotically anti--de Sitter case,
  the stability analysis requires only that the black hole have a
  single Killing field normal to the horizon and there are no
  restrictions on the perturbations (apart from smoothness and
  appropriate behavior at infinity).  For an asymptotically anti--de
  Sitter black hole, we define an {\em ergoregion} to be a region
  where the horizon Killing field is spacelike; such a region, if
  present, would normally occur near infinity.  We show that for black
  holes with ergoregions, initial data can be constructed such that
  $\mathcal{E}<0$, so all such black holes are unstable. To obtain
  such initial data, we first construct an approximate solution to the
  constraint equations using the WKB method, and then we use the
  Corvino-Schoen technique to obtain an exact solution. We also
  discuss the case of charged asymptotically anti--de Sitter black
  holes with generalized ergoregions.
\end{abstract}

\maketitle

\section{Introduction}

If a suitably tuned wave impinges upon a rotating object, then the
amplitude of the reflected wave exceeds that of the incident wave---a
phenomenon known as
superradiance~\cite{Zeldovich:1972,Starobinskii:1973}.  Rotating black
holes with ergoregions are examples of such systems, making it
possible to extract energy from them~\cite{Penrose:1969pc}. It is
intuitively clear that if one were to surround such a black hole by a
suitable mirror that is far enough away, then the amplitude of the field representing the wave
would grow unboundedly due to repeated superradiant scattering.
In~\cite{Press:1972zz} the authors investigated how to effectively
make such mirrors by appropriate matter fields, coining the
terminology ``black hole bomb'' for the resulting instability.

With the advent of the AdS-CFT correspondence~\cite{Maldacena:1997re},
it was soon realized that AdS boundary conditions are an alternative
way to produce a mirror~\cite{Hawking:1999dp}, and can therefore lead
to superradiant instabilities. For instance, sufficiently small Kerr-AdS black
holes (below the Hawking-Reall bound~\cite{Hawking:1999dp}) have
ergoregions, and have indeed been shown to be unstable to scalar field
perturbations~\cite{Cardoso:2004hs,Dold:2015cqa}.

An even more interesting possibility is that the black hole can be
unstable to perturbations of the gravitational field itself.  The
standard approach to identify such a (linear) instability is to search
for mode solutions that grow in time, which requires solving the
linearized Einstein equations in time. For the Kerr-AdS metric in 4
dimensions the linearized equations can be decoupled and separated
into modes using the Teukolsky method~\cite{Teukolsky:1973ha}, making
such an analysis feasible in principle, but very difficult in
practice~\cite{Cardoso:2013pza}.  However, in higher dimensions or in more
complicated backgrounds (e.g., with other matter fields or less symmetry), this method fails.  
For this reason, there has been limited
success in demonstrating that the expected superradiant instability
actually occurs in general.

In this paper, we will use an alternative method to show the
occurrence of an instability associated with superradiance of
gravitational perturbations for a very wide class of asymptotically
AdS black holes.  Our approach is based on the so-called ``canonical
energy method\footnote{The method was used by
  Friedman~\cite{Friedman:1978} to study the stability of relativistic
  stars.}''~\cite{Hollands:2012sf}.  The canonical energy
$\mathcal{E}(\gamma)$ is an integral over a Cauchy hypersurface
$\Sigma$ of the region exterior to the black hole, quadratic in the
perturbation $\gamma_{ab}$.  $\mathcal{E}(\gamma)$ can be written in
terms of the initial data on $\Sigma$ of $\gamma_{ab}$, so in practice
we must only solve the linearized constraint equations for the initial
data on $\Sigma$, rather than the full evolution equations in $M$. As
a consequence, the analysis is greatly simplified compared to the
standard approach. In the asymptotically flat case, the canonical
energy can be proven to be gauge invariant for perturbations that fix
the black hole area and the linear momentum, and for perturbations
that also fix the mass and angular momenta, $\E(\gamma)$ can be proven
to be degenerate if and only if $\gamma_{ab}$ is a perturbation to
another stationary black hole. (The corresponding results in
asymptotically AdS spacetimes will be given in
lemmas~\ref{lemma:gaugeinv} and \ref{nondeglemma} below.)
Furthermore, the value of $\E(\gamma)$ is independent of the choice of
Cauchy surface $\Sigma$, but, for axisymmetric perturbations, its flux
through the horizon and infinity is positive, so it decreases in time
in the sense that its value on a slice $\Sigma'$ that terminates at
the future horizon and/or future null infinity is smaller (see
\fref{DOC2} below for such a slice in the asymptotically AdS case).

These properties  give rise to the following stability
criterion: If $\E(\gamma)$ is non-negative on a
space of perturbations that fix appropriate conserved quantities, then it is positive definite 
on this space
modulo perturbations to other stationary black holes, implying mode stability\footnote{In the terminology of 
dynamical systems, it implies ``orbital stability''.}. 
Conversely, if for some $\gamma_{ab}$ in this space we have
$\mathcal{E}(\gamma) < 0$, then since the canonical energy can
only decrease in time, $\gamma_{ab}$ cannot settle down to a
stationary configuration (since the canonical energy vanishes
for stationary perturbations in this space). Thus, it corresponds to an instability. Furthermore, 
for axisymmetric perturbations, one can prove that the ``kinetic energy'' 
is always positive, thereby enabling one to obtain 
results~\cite{Prabhu:2015rua} on exponential growth of perturbations
when $\mathcal{E}(\gamma) < 0$. The problem of establishing 
the existence of a perturbation 
$\gamma_{ab}$ with negative canonical energy is, of course, a much simpler problem than
that of solving the evolution equations for gravitational perturbations. 

In the case of a black hole in an asymptotically flat spacetime, we have to consider the flux
of canonical energy at both the event horizon, $\eH^+$, and at 
null infinity, $\scri^+$. As shown in~\cite{Hollands:2012sf}, if the canonical energy 
$\E = \E_K$ is defined with respect to the horizon Killing field $K^a$,  
then the flux of canonical energy is positive at $\eH^+$. On the other hand, if 
$\E = \E_T$ is defined  with respect to the stationary Killing field $T^a$, (i.e., the Killing
field that is timelike near infinity) then the flux is positive at $\scri^+$. Consequently, for a rotating black hole,
one must restrict to axisymmetric perturbations in order
that the two canonical energies coincide, $\E_K = \E_T$, so that one has a positive flux at
both boundaries, thereby enabling one to prove stability and instability results~\cite{Hollands:2012sf}. 
 
In the case of a stationary relativistic star in an asymptotically
flat spacetime, there is only one boundary (namely, $\scri^+$) through
which there can be a flux of canonical energy. Consequently, one can
work with the canonical energy, $\E_T$, defined with respect to the
stationary Killing field, $T^a$, and there is no need to restrict to
axisymmetric perturbations~\cite{Friedman:1978}.  However, the
exponential growth results of~\cite{Prabhu:2015rua} do not apply to
non-axisymmetric perturbations, since the ``kinetic energy'' need not
be positive for non-axisymmetric perturbations, i.e., if $\E_T < 0$
for a non-axisymmetric perturbation of a rotating star, one can prove
instability only in the sense that this perturbation cannot
asymptotically approach a stationary perturbation. 

A similar situation occurs for the case of interest here, namely, a
black hole in an asymptotically AdS spacetime possessing a Killing
field, $K^a$, normal to the horizon. In this case, there is again only
one boundary through which there can be a flux of canonical energy,
but now this boundary is $\eH^+$ rather than $\scri$. Consequently,
one can now work with the canonical energy $\E_K$, and there is no
need to restrict to axisymmetric perturbations. However, again the
exponential growth results of~\cite{Prabhu:2015rua} do not apply to
non-axisymmetric perturbations\footnote{It is thus hard to
  characterize by our method the actual nature of the
  instability. However, given that generically one has a flux of
  positive energy through the horizon, it is very implausible to
  imagine a mechanism by which the perturbation could actually fail to
  blow up (making $\E_K$ tend to minus infinity). In particular, one ought
  to be able to rule out a scenario wherein the perturbation
  approaches a time-periodic solution (i.e. one for which there is an
  isometry $\varphi$ of the background such that $\varphi(x) \in
  J^+(x)$ for all $x \in M$ and $\varphi^* \gamma_{ab} =
  \gamma_{ab}$.) Such a condition implies vanishing flux through the
  horizon and hence vanishing (perturbed) expansion, shear and
  twist. By results of~\cite{Moncrief:1983xua,Moncrief:2008mr}, this
  would imply (for analytic $\gamma_{ab}$) that $\gamma_{ab}$ is
  actually a perturbation towards another stationary black hole, a
  contradiction.  }.

Consider a test particle of $4$-momentum $p^a$ 
propagating in an asymptotically AdS spacetime containing a black
hole with horizon Killing field $K^a$. The energy of the particle with respect to $K^a$ is given by
\ben
\E_{K, \rm particle} = -K^a p_a  \ . 
\een
Thus, if $p^a$ is (future-directed) timelike or null, then $\E_{K, \rm particle} > 0$ everywhere
that $K^a$ is (future-directed) timelike. Conversely, if there is an
{\em ergoregion} in the spacetime---i.e., a region where $K^a$ is spacelike---then $\E_{K, \rm particle}$
can be made negative by suitably choosing a timelike or null $p^a$ in the ergoregion. This suggests
that if an ergoregion is present, we should be able to find a gravitational perturbation for which the canonical 
energy defined with respect to $K^a$ satisfies
$\mathcal{E}_K(\gamma) < 0$. 
If so, by the above argument, the black hole would then be 
unstable to gravitational perturbations. However,
it is not obvious that we can find a gravitational perturbation with $\mathcal{E}_K(\gamma) < 0$ whenever an
ergoregion is present because (i) the gravitational perturbation is a wave, not a particle, and 
is therefore ``spread out'' and (ii) we must
ensure that the constraint equations hold for the initial data for this wave, so we are 
restricted in the initial conditions we can choose. 

The main result of this paper is the following theorem:

\begin{thm} \label{thm1}
Let $(M,g_{ab})$ be a $d$-dimensional $(d \ge 4)$ asymptotically AdS black hole with Killing horizon
and corresponding Killing field $K^a$ (but not necessarily any further ones). 
If $K^a$ is spacelike at some point $x$
in the domain of outer communication, $\M$, then there exists a perturbation $\gamma_{ab}$
with compactly supported initial data near $x$ such that $\mathcal{E}_K (\gamma)<0$.  
In particular, the perturbation cannot settle down to a perturbation to another stationary black hole, showing that 
the black hole is unstable (in this sense). 
\end{thm}

To construct the desired perturbation $\gamma_{ab}$ with $\mathcal{E}_K (\gamma)<0$, 
it is natural to seek $\gamma_{ab}$ in the form of a high frequency gravitational wave 
describing a null particle with momentum $p^a$ in the optical approximation, 
and it is natural to expect that, for 
such a wave, $\E_K (\gamma) \propto \E_{K, \rm particle}$ approximately. This expectation turns out to be broadly correct, but 
the precise argument is complicated by the fact that the high frequency ansatz only gives an approximate solution.
 
To address this issue, our proof proceeds in two steps. 
\begin{enumerate}
\item We use the WKB method to obtain an approximate solution of the form 
$\gamma_{ab} = A_{ab} \exp(i\omega \chi)$, where $\omega \gg 1$ is the WKB frequency 
parameter and $\chi$ is a phase function satisfying the usual  eikonal equation  $p^a p_a = 0$,
where $p_a = \nabla_a \chi$ is interpreted as the momentum 
of the high frequency gravitational wave. $A_{ab}$ is an amplitude satisfying transport equations, 
which is chosen to be of compact spatial support and
sharply peaked in a small neighborhood of a particle trajectory with tangent $p^a$
going through a point $x$ where $K^a$ is space like. As in the particle case, we choose $p^a$ 
such that $K^a p_a > 0$ near this trajectory. 
It will then be seen that  $\mathcal{E}_K(\gamma) \sim -\omega^2 K^a p_a  < 0$ for 
an appropriate surface $\Sigma$ containing $x$, see equation~\eref{Eineq} and footnote~\ref{footnote6}. In other words, 
the canonical energy is approximately equal to the energy of a particle with momentum $p_a$ 
relative to the Killing field $K^a$. 

\item The WKB solution is not an exact solution, and so its initial data do not satisfy the 
linearized constraints. But we can, following~\cite{Hollands:2014lra}, 
fix this up by an application of the Corvino-Schoen
  method~\cite{Corvino:2003sp,Chrusciel:2003sr}.  For $\omega \gg 1$, 
  the correction is small and it remains true that $\mathcal{E}_K(\gamma)<0$ for the 
  corrected perturbation.
\end{enumerate}

We finally note that the statement $\mathcal{E}_K(\gamma)<0$ of our
theorem would actually also hold for a black hole in an asymptotically
flat spacetime (with identical proof) whenever there is a region where
$K^a$ is spacelike. Similarly, by the same construction, we could
obtain another perturbation ${\gamma}_{ab}$ with $\E_T(\gamma) < 0$
whenever $T^a$ is spacelike.  However these results do {\em not} imply
a superradiant instability in the asymptotically flat case because, as
discussed above, the flux of $\E_K$ through $\scri^+$ need not be
positive for non-axisymmetric perturbations and, similarly, the flux
of $\E_T$ through $\eH^+$ need not be positive for non-axisymmetric
perurbations.  Thus, the instability statement our theorem applies
only to the asymptotically AdS case\footnote{ Note, however, that our
  theorem does apply to the case of ``black hole bomb,'' in which the
  spacetime is vacuum with no negative cosmological constant but the
  black hole is surrounded by some effective mirror that reflects all
  perturbations.  The mirror would have to be placed far enough from
  the black hole that an ergoregion with respect to the horizon
  Killing field, $K^a$, is present}.

The plan of this paper is as follows: In \sref{can}, we elaborate on
our assumptions about the backgrounds, and review the canonical energy
method~\cite{Hollands:2012sf}, adapted to the asymptotically AdS
case~\cite{Hollands:2014lra}. In \sref{wkb}, we construct the WKB
``solutions''. In \sref{neg} we show how to correct them and complete
the proof of theorem~\ref{thm1}, thereby establishing the superradiant
instability.

\medskip
\noindent \textbf{Notations and conventions:} We follow the conventions
of~\cite{Wald:1984}.

\section{Canonical energy and stability in asymptotically
  AdS spacetimes}\label{can}

The backgrounds we consider are asymptotically $d$-dimensional
$(d\ge 4)$ AdS black hole spacetimes $(M,g_{ab})$ with 
Killing horizon. The horizon Killing field is denoted $K^a$. As usual, on the horizon
it satisfies
\ben
K^b \nabla_b K^a = \kappa K^a \ , 
\een
where $\kappa \ge 0$ is the surface gravity (see sec.~12.5 of~\cite{Wald:1984}). 
The spacetime is called degenerate
(extremal black hole) if $\kappa = 0$; otherwise it is called non-degenerate. 

The precise asymptotic conditions are formulated within the standard
framework of conformal infinity (see, e.g., section 11.1
of~\cite{Wald:1984}): There should exist a conformal completion
$(\tilde M, \tilde g_{ab}=\Omega^2 g_{ab})$ such that $\Omega$
vanishes on the conformal boundary $\scri = \partial \tilde M$, and
such that $\tilde g_{ab}$ is smooth across $\scri$. Throughout we
assume the Einstein equations $G_{ab} = -\Lambda g_{ab}$ with negative
cosmological constant $\Lambda<0$. As is well-known, this implies that
$\scri$ is timelike. We require the strengthened global AdS-type
boundary condition that $\scri \cong \RR \times S^{d-2}$ topologically
and metrically, i.e.,  that the induced metric $h_{ab}$ on $\scri$ is
$\ell^2$ times that of the Einstein static universe, where 
$\ell = \sqrt{-(d-1)(d-2)/2\Lambda}$ is the AdS-radius. It is standard
to show that these assumptions imply the asymptotic expansion
\ben\label{bndy}
 g_{ab} = \Omega^{-2}\left( h_{ab} + \Omega^{d-1}
  E_{ab} + O(\Omega^{d}) \right) \ 
  \een 
  for a suitable choice of
$\Omega$ (``Graham-Fefferman gauge'') assumed from now on.  $\nabla_a
\Omega$ is normal to $\scri$ in the sense that $h^{ab} \nabla_a \Omega
= 0$ on $\scri$, and $E_{ab}$ is intrinsic to $\scri$ in the sense
that $E_{ab} \nabla^a \Omega = 0$, as well as being transverse and
traceless\footnote{The tensor turns out to be equal to the limit
  $E_{ab} = \lim_{\scri} \frac{1}{d-3} \Omega^{-d+1} C_{acbd}
  (\nabla^c \Omega) \nabla^d \Omega$ at $\scri$.}.  In~\eref{bndy}, and
in the following, the ``big-O'' notation $O(\Omega^n)$ means a
function on $\tilde M$ such that $\Omega^{-n} O(\Omega^{n})$ is
smooth at $\scri$. For details, see, e.g.,~\cite{Hollands:2005wt}.

The domain of outer communication is $\M \equiv J^+(\scri) \cap J^-(\scri)$.
The inner
boundaries of $\M$ are then by definition the future and past horizons
$\eH^\pm = \partial \M \cap J^\mp(\scri)$. We demand that these horizons
be Killing horizons, i.e., there exists a Killing field $K^a$ that
is tangent to the generators of $\eH^\pm$.  Examples of such
spacetimes are provided by the AdS-Myers-Perry
metrics~\cite{Myers:1986un,Gibbons:2004js}. These have additional Killing
fields beyond $K^a$, and they are in particular stationary. But our analysis will not require any of these
and just use $K^a$. In particular, there is numerical
evidence~\cite{Dias:2015rxy,Niehoff:2015oga} for the existence of
asymptotically AdS black holes where there are no additional Killing
fields besides $K^a$, and our results would apply to these black holes.
We will, by a slight abuse of terminology, still refer to our black holes as ``stationary''. 

Linear perturbations are solutions $\gamma_{ab}$ to the linearized
Einstein equations \ben\label{ein} 0=\nabla^b \nabla_b \gamma_{ac} -2
\nabla_{(a} \gamma_{c)} + 2R_a{}^b{}_c{}^d\gamma_{bd} , \een where
$\gamma_c = \nabla^b(\gamma_{cb} - \frac{1}{2} g_{cb} \gamma)$, where
$\gamma = \gamma_a{}^a$, and where indices are raised with $g^{ab}$.
Under the imposition of the gauge condition\footnote{\label{gaugef}
  The fact that we can impose this gauge follows as usual from the
  fact that $\gamma_c \to \gamma_c + \nabla^a\nabla_a v_c +
  \frac{2}{d-2} \Lambda v_c$ under the infinitesimal gauge
  transformation $\gamma_{ab} \to \gamma_{ab} + \pounds_v g_{ab}$. The
  residual gauge freedom consists of vector fields $v^a$ satisfying
  $\nabla^a\nabla_a v_c + \frac{2}{d-2} \Lambda v_c=0$. It may be used
  to additionally impose $\gamma = 0$ as a gauge condition since
  \eref{ein} and $\gamma_c=0$ implies the wave equation $\nabla^a
  \nabla_a \gamma + \frac{4}{d-2}\Lambda \gamma=0$ for the trace.
  Indeed, choosing initial data for $v^a$ on some Cauchy-surface
  $\Sigma$ such that $\gamma=0, n^a \nabla_a \gamma = 0$ we obtain
  $\gamma =0$ in the domain of dependence; see sec.~7.5
  of~\cite{Wald:1984} for details.} $ \gamma_c =0$ this equation takes
the form of a standard wave equation. Hence, it has a well-posed
initial value formulation in any globally hyperbolic subset of $\M$.
As usual, due to the presence of the time-like (conformal) boundary,
the entire domain of outer communication $\M$ is not globally
hyperbolic.  To get a well-defined initial value problem throughout
$\M$, one has to impose boundary conditions at $\scri$. We impose the
``reflecting'' boundary conditions given by the linearization
of~\eref{bndy}, i.e., we postulate $\gamma_{ab} = O(\Omega^{d-3})$ near
$\scri$. Under these conditions, it can be shown that any smooth
initial data $(\delta q_{ab}, \delta p^{ab})$ for $\gamma_{ab}$
satisfying the linearized constraints (see below) as well as $\delta
q_{ab} = O(\Omega^{d-3}), \delta p_{ab} = O(\Omega^{d-4})$ lead to a
unique smooth, globally defined solution $\gamma_{ab}$ satisfying the
gauge condition throughout $\M$~\cite{Friedrich:1995vb}. Here, the
initial data is to be specified on a smooth, acausal hypersurface
$\Sigma$ such that every inextendible timelike curve in the domain of
outer communications that does not have an endpoint on $\scri$ must
intersect $\Sigma$. By another slight abuse of terminology,
we will refer to a hypersurface $\Sigma$ with this
property as a {\em Cauchy surface} for $\M$, even though, of course,
asymptotically AdS spacetimes are not globally hyperbolic and do not
admit a Cauchy surface in the usual sense.

Given a pair of perturbations $\gamma_{1ab}, \gamma_{2ab}$ satisfying \eref{ein}, one defines the ``symplectic current'' by 
\begin{equation}
  w^a (\gamma_1, \gamma_2) = \frac{1}{16\pi} g^{abcdef}\left(\gamma_{2bc} \nabla_d\gamma_{1 ef} - \gamma_{1bc}\nabla_d\gamma_{2 ef} \right),
  \label{w}
\end{equation}
where
\begin{equation}
 g^{abcdef} = g^{ae}g^{fb}g^{cd} - \frac{1}{2}g^{ad}g^{be}g^{fc} - \frac{1}{2} g^{ab} g^{cd} g^{ef} -\frac{1}{2}g^{bc}g^{ae}g^{fd} + \frac{1}{2}g^{bc}g^{ad}g^{ef}.
\end{equation}
The symplectic current is shown to be conserved, $\nabla_a w^a = 0$, and its existence is best understood from the variational principle 
underlying the Einstein equations (see~\cite{Hollands:2012sf}). The ``symplectic form'' is then
obtained by integrating $w^a$ over a Cauchy surface  $\Sigma$ (in the sense defined at the end of
the previous paragraph),
\begin{equation}\label{symform}
 W_\Sigma(g;\gamma_1,\gamma_2) = \int_\Sigma n_a w^a (\gamma_1, \gamma_2)  \ , 
\end{equation}
where $n^a$ is the future-directed time like normal to $\Sigma$, and
where the natural volume element on $\Sigma$ is understood.  

\begin{figure}[t]
\begin{center}
\begin{tikzpicture}[scale=1, transform shape]
\filldraw[fill=gray,opacity=.6,draw=black] (-2,0) -- (0,0) -- (0,1) -- (-1,1) -- (-2,0);
\draw[double] (0,2) --  (0,-2);
\draw (0,2) -- (-2,0) -- (0,-2);
\draw[red,thick] (-2,0) -- (0,0);
\node at (-1,-.4) {$\Sigma$};
\draw[red,thick] (-1,1) -- (0,1);
\node at (-.5,.6) {$\Sigma'$};
\draw (-2,0) node[draw,shape=circle,scale=0.3,fill=black]{};
\draw (0,2) node[draw,shape=circle,scale=0.3,fill=black]{};
\draw (0,-2) node[draw,shape=circle,scale=0.3,fill=black]{};
\node at (-2.1,.5) {$\mathscr{H}_{12}$};
\node at (.4,.5) {$\mathscr{I}_{12}$};
\node at (-1.6,1) {$\mathscr{B}'$};
\node at (-2.6,0) {$\mathscr{B}$};
\node at (.4,1) {$\mathscr{C}'$};
\node at (.4,0) {$\mathscr{C}$};
\draw (-1,1) node[draw,shape=circle,scale=0.3,fill=red]{};
\draw (0,1) node[draw,shape=circle,scale=0.3,fill=red]{};
\end{tikzpicture}
\end{center}
\caption{
\label{DOC2}
Conformal diagram of the exterior of the $AdS$ black hole.  To obtain
the balance equation, we integrate $\nabla^a w_a = 0$ over the shaded
rectangle. In this case, there is no flux across $\mathscr{I}_{12}$
due to the AdS boundary conditions.  }
\end{figure}
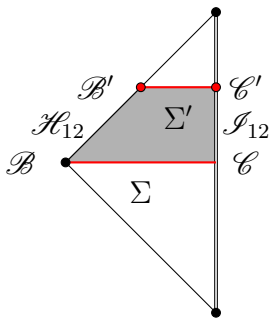

The canonical energy $\E_K$ with respect to the horizon Killing field $K^a$ of
a perturbation $\gamma_{ab}$ 
is simply the symplectic product of $\gamma_{ab}$ with its ``time derivative''
$\pounds_K \gamma_{ab}$,
\begin{equation}
 \E_K (\gamma) = W_\Sigma(g; \gamma, \pounds_K \gamma) \ , 
\end{equation}
However, as defined, $\E_K$ is not gauge invariant. To obtain a gauge invariant
quantity, we need to fix  the gauge at
the horizon and at $\scri$. In addition, we wish to define canonical energy
not only on Cauchy surfaces $\Sigma$ but also on slices $\Sigma'$ that 
extend from $\mathscr{H}^+$ to $\scri$ (see \fref{DOC2}).
In order to obtain a quantity with good monotonicity properties, it is useful to 
introduce an additional boundary term into the definition of canonical energy
when evaluated on such slices. We now state the gauge conditions that we impose and
we then introduce this boundary term.

The gauge conditions are as follows: Near $\scri$ we impose on $\gamma_{ab}$ the linearized
version of the Graham-Fefferman type gauge~\eref{bndy}, meaning that
\ben 
\Omega^{-d+1} \gamma_{ab} \nabla^a \Omega = 0 , \qquad
\Omega^{-d+1} \gamma = 0 , \qquad D^a(\Omega^{-d+1} \gamma_{ab}) = 0
\een 
in the limit at $\scri$. These conditions imply in particular the
convergence of~\eref{symform} (see,
e.g.,~\cite{Hollands:2005wt}). Near $\eH^+$, we can first impose the
linearized ``Gaussian normal null form'' gauge conditions described
in~\cite{Hollands:2012sf}.  As in that reference, we would
additionally like to impose the conditions that the perturbed
expansion\footnote{Here we use the standard convention that $\delta X$
  denotes the first order perturbation of a quantity $X$. More
  precisely, if $g_{ab}(\lambda)$ is a differentiable 1-parameter
  family of metrics with
  $\gamma_{ab} = dg_{ab}(\lambda)/d\lambda |_{\lambda = 0}$, and if
  $X$ depends on $g_{ab}$ in a differentiable manner, then
  $\delta X = d X(g(\lambda))/d\lambda |_{\lambda = 0}$.},
$\delta \vartheta$, and area element, $\delta \epsilon$, of $\gamma_{ab}$, vanish on
$\eH^+$. In~\cite{Hollands:2012sf} a proof was given that a choice of gauge can always
be made to impose $\delta \vartheta = 0$ for
non-degenerate Killing horizons. This proof relies crucially on a stability 
property of marginally outer trapped surfaces and does not generalize straightforwardly to
degenerate Killing horizons. However, 
we shall be interested here only in perturbations 
with initial data of compact support (away from the horizon)
on a Cauchy surface $\Sigma$.
It is easily seen that the desired gauge condition
automatically holds for all such perturbations by the following argument:
By the compact support and domain of dependence property
$\delta \vartheta$ vanishes for sufficiently negative values of the affine
parameter $u$ on $\eH^+$. However, by the linearized
Raychaudhuri equation, we have
\ben
\label{ray} 
\frac{\D }{\D u}
\delta \vartheta = - \frac{2}{d-2} \vartheta \delta \vartheta -
2\sigma_{ab} \delta \sigma^{ab} - \delta R_{ab} k^a k^b = 0 \ , 
\een
where $k^a=(\partial/\partial u)^a$ is the affinely parametrized tangent to the null generators of $\eH^+$, and
where $\sigma_{ab}$ and $\vartheta$ are the (vanishing) shear and
expansion of the background and $\delta \sigma_{ab}$ and
$\delta \vartheta$ are their first order variation under $\gamma_{ab}$.
Thus, $\delta \vartheta$ vanishes everywhere on $\eH^+$, as desired.
It then also follows that the perturbed area element, $\delta \epsilon |_{\eB}$,
vanishes on any cross section $\eB \subset \eH^+$, so we have the desired conditions
\ben\label{horgauge} 
\delta \epsilon |_{\eB} = 0 = \delta
\vartheta|_\eB \ .
\een 
As already stated, for the non-degenerate case, the condition $\delta \vartheta = 0$ can be imposed
by a gauge choice without assuming that the perturbation is initially supported away from the horizon.
In addition, it is easily seen~\cite{Hollands:2012sf} that the 
condition $\delta \epsilon = 0$ can be imposed by a choice 
of gauge provided only that $\delta A = 0$ at the initial time. 

We now define a boundary term, $B_\eB(g; \gamma)$,
associated with a cross-section $\eB$ of the future horizon by
\ben\label{Bgammadef} 
B_\eB(g; \gamma) = \frac{1}{32\pi}
\int_\eB \gamma^{ab} \pounds_K \gamma_{ab} \ .  
\een 
Here,  the area element $\epsilon |_{\eB}$ is
understood in the integral. 
For a slice, $\Sigma'$, that extends
from an arbitrary cross section $\eB$ of $\eH^+$ to a cross section $\eC$ of $\scri$,
we define the canonical energy (with boundary term) by\footnote{The corresponding quantity
in the asymptotically flat case was called the ``modified canonical energy'' in \cite{Hollands:2012sf}.}
\ben\label{Edef}
\overline \E_K(\gamma, \Sigma') \equiv \int_{\Sigma'} n_a w^a(\gamma, \pounds_K \gamma) - B_\eB(g; \gamma)   . 
\een
where $w^a$ was defined by \eref{w}. Here we put an overline on $\overline \E_K$ to emphasize that
we are allowing $\Sigma'$ to be an arbitrary slice as depicted in \fref{DOC2}. 
In the non-degenerate case, if we evaluate $\overline \E_K$ on a Cauchy surface $\Sigma$, then 
$\Sigma$ will extend to the bifurcation surface, where $K^a = 0$. 
Hence $B_{\eB}(g; \gamma) = 0$, so for any Cauchy surface $\Sigma$, we have
\ben
\overline \E_K(\gamma, \Sigma) = W_\Sigma(g; \gamma, \pounds_K \gamma) \equiv \E_K(\gamma, \Sigma) \ .
\een 
Similarly, in the degenerate case, we have $\overline \E_K(\gamma, \Sigma) = \E_K(\gamma, \Sigma)$
for any Cauchy surface $\Sigma$ if the initial data for $\gamma_{ab}$ is of compact support on $\Sigma$.

The reason for introducing the boundary term in the definition of
$\overline \E_K$ is a very important \emph{monotonicity property} under `time
evolution'. This property is obtained as follows. We integrate the
equation $\nabla_a w^a (\gamma, \pounds_K \gamma) = 0$ over a quadrangle-shaped domain of $\M$
 bounded by a Cauchy surface $\Sigma$ and a slice $\Sigma'$ 
as shown in \fref{DOC2}. 
By Stokes' theorem, the result is a contribution from the
boundaries. The contributions from $\Sigma$ is $\E_K$. There is no contribution
from $\scri_{12}$. By the same calculation as done 
in~\cite{Hollands:2012sf} and~\cite{Hollands:2014lra}
the contribution from $\eH_{12}$ is positive up to the boundary term
\eref{Bgammadef}. By incorporating this boundary term into the definition
of $\overline \E_{K}(\gamma, \Sigma')$, we obtain the following result

\begin{lemma}\label{fluxlemma}
  Let $\Sigma$ be a Cauchy surface for the domain of outer communications and
  let $\Sigma' \subset J^+(\Sigma)$, as depicted in
  \fref{DOC2}. Then for any perturbation $\gamma_{ab}$ (assumed to be of compact support
  on $\Sigma$ in the degenerate case) we have
     \ben 
     \overline \E_K(\gamma, \Sigma) -
  \overline \E_{K}(\gamma, \Sigma') = \frac{1}{4\pi} \int_{\eH_{12}} (K^c\nabla_c
  u) \delta \sigma_{ab} \delta \sigma^{ab} \ge 0 \ , \een so that
  $\overline \E_{K} (\Sigma')  \le  \E_{K}$ ($u$ is the affine parameter along the horizon used in the definition of
the shear, $\sigma_{ab}$).
\end{lemma}

The monotonicity of $\overline \E_K$ expressed by the lemma is the
first main ingredient in the (in-)stability argument.  The second
important ingredient is an analysis of the subspace of perturbations where 
the canonical energy is (non-) degenerate~\cite{Hollands:2012sf}. To formulate
it, recall that an asymptotic symmetry $X^a$ is a vector field that is
smooth on $\tilde \M$ such that its restriction is tangent to $\scri$
and defines a conformal Killing field of the induced metric $h_{ab}$
on $\scri = \RR \times S^{d-2}$. Two asymptotic symmetries are
considered equivalent if their restrictions to $\scri$ coincide. The
ADM-type conserved quantity associated with such an equivalence class
$X^a$ is~\cite{Hollands:2005wt} 
\ben\label{HXdef} 
H_X =
-\frac{\ell}{8\pi} \int_{\eC} E_{ab} X^a \tilde n^b \ , 
\een 
where $\eC \cong S^{d-2}$ is an arbitrary cut of $\scri$ (see \fref{DOC2}), where the integration 
element induced from $h_{ab}$ is
understood, and where $\tilde n^b$ is the timelike normal to 
$\eC$ within $\scri$ (relative to $h_{ab}$). Since the algebra of conformal Killing fields (CKVs) on
$\RR \times S^{d-2}$ is $\so(2,d-1)$, it follows that the asymptotic
symmetries modulo equivalence are in one-to-one correspondence with
the generators of this Lie algebra.

Canonical representers of a natural basis of $\so(2,d-1)$ in
``AdS-embedding coordinates'' are given in eq.~\eref{generators} in
\ref{appendix}. They include a globally timelike asymptotic symmetry
$T^a$ and asymptotic symmetries $\psi_{ij}^a, 1\le i<j \le d-1$ with
closed orbits corresponding to (asymptotic) rotations in the
``$ij$-plane''.  Their restrictions to $\scri$, relevant
for~\eref{HXdef} may also be computed. For instance, if we choose
coordinates $(t,\mu_I,\phi_I)$ on $\RR \times S^{d-2}$ such that the
induced metric reads\footnote[1]{Here we assume that $d$ is odd for
  definiteness.}  $h = \ell^2[-dt^2 + \sum (d\mu_I^2 + \mu_I^2
d\phi_I^2) + dz^2]$, where $I=1, \dots, \lfloor \half (d-3) \rfloor, z
= (1-\sum \mu_I^2)^\half$, then \ben T^a = \left(
  \frac{\partial}{\partial t} \right)^a \ , \quad \psi_{12}^a = \left(
  \frac{\partial}{\partial \phi_1} \right)^a \ , \quad \psi_{34}^a =
\left( \frac{\partial}{\partial \phi_2} \right)^a \ , \quad \dots \ ,
\een for further details see~\cite{Chrusciel:2006zs,Chen:2015gav}.
The conserved quantity associated with $T^a$ is the mass $m=H_T$, the
conserved quantity associated with $\psi_{ij}^a$ the angular momentum
$J_{ij} = H_{\psi_{ij}}$ in the $ij$-plane, etc.

We are now in a position to characterize the gauge
invariance of the canonical energy. Our gauge conditions 
imposed near the horizon and infinity leave us with the freedom of applying the remaining ``admissible'' 
gauge transformations $\gamma_{ab} \to \gamma_{ab} + \pounds_X g_{ab}$, where $X^a$ must satisfy:
\ben\label{gaugerem} 
X^a = 
\Bigg\{
\begin{array}{cc}
fk^a  &  {\rm on} \ \eH^+, \ {\rm with} \ k^a \nabla_a f= 0, \\
{\rm conformal \ Killing \ vector \ field} & {\rm on} \ \scri, \ {\rm with} \  X^a \nabla_a \Omega=0  \ .
\end{array}
\een 
The subspace of (smooth) perturbations on which the canonical energy is gauge 
invariant for all such $X^a$ is given by the following lemma:

\begin{lemma}\label{lemma:gaugeinv}
Let $\Sigma$ be a Cauchy surface for the domain of outer communications. The canonical energy $\E_K(\gamma, \Sigma)$ 
is invariant under $\gamma_{ab} \to \gamma_{ab} + \pounds_X g_{ab}$ for all admissible $X^a$
(i.e., those satisfying \eref{gaugerem}) 
precisely on the space of smooth perturbations 
$\gamma_{ab}$ solving the linearized Einstein equations, 
satisfying our gauge conditions \eref{horgauge}, 
and satisfying $\delta H_Y = 0$ for all asymptotic symmetries 
$Y \in \g_K \equiv \{ Y = [K,Z] \mid Z \in \so(2,d-1)\}$. 
\end{lemma}

The proof of this lemma is completely analogous to that of
proposition 3 of~\cite{Hollands:2012sf} in the
asymptotically flat case. The only difference lies in the fact that,
in the asymptotically flat case, we were interested in $\E_T$, and
that the algebra of asymptotic symmetries is isomorphic to the
Poincare Lie algebra, $\so(1,d-1) \ltimes \RR^{d}$, rather than
$\so(2,d-1)$. Consequently, in the asymptotically flat case, the gauge
invariance would now hold for perturbations $\gamma_{ab}$ so that
$\delta H_Y = 0$ where $Y$ is now an element of $\{Y = [T,Z]
\mid Z \in \so(1,d-1) \ltimes \RR^d\}$. It is not hard to see that, in
the asymptotically flat case, this means concretely that $\delta p_i = 0$, where $p_i =H_{\partial/\partial x^i} , i=1, \dots, d-1$ are the
components of the ADM (linear) momentum.  In the asymptotically AdS case, we need
to consider instead $\g_K$, and this subspace depends strongly on the
actual form of $K^a$ near $\scri$.  We show in~\ref{appendix} that
$K^a$ can be brought into one of the canonical forms (s0)--(s3), (n1),
(n2) presented in lemma~\ref{normformlemma} (table~\ref{tab:table1})
near $\scri$ by a diffeomorphism representing an asymptotic symmetry.
It is thus enough to calculate $\g_K$ for each of these normal
forms. As an example, consider the normal form (s0) in $d=4$, which
means that, near $\scri$, we have \ben K^a = T^a + \Omega \psi_{12}^a
\ , \een where $\Omega \in \RR$ is the angular velocity of the horizon
(not to be confused with the conformal factor). If we also assume that
$\Omega \neq 0$, then $\g_K$ is spanned by the asymptotic symmetries
$\psi_{13}^a, \psi_{23}^a$ together with $ P_1^a, P_2^a, P_3^a$ and
$C^a_1, C_2, C^a_3$, see~\eref{generators}. Thus, gauge invariance
requires $\delta J_{13} = \delta J_{23} = 0$ as well as $\delta p_i =
\delta c_i = 0$ where $p_i = H_{P_i}, c_i = H_{P_i}$. Different
conditions would be obtained e.g., if $\Omega=0$, or if $K^a$ has
another one of the normal forms. \qed

To fully characterize the degeneracies of $\E_K$, we need the notion of a ``perturbation 
towards another stationary black hole''. Following~\cite{Hollands:2012sf}, this notion is defined as follows. 
Let $g_{ab}(\lambda)$ be a 1-parameter family of asymptotically AdS metrics with Killing horizon
described above, which for $\lambda=0$ coincides with our given background, $g_{ab}=g_{ab}(0)$. 
Let $K^a(\lambda)$ be the horizon Killing field and $\kappa(\lambda)$ the surface gravity. 
We can make a gauge choice so that, near $\eH^+$, we have $K^a(\lambda) = (\kappa(\lambda)/\kappa) K^a$, whereas near $\scri$, 
we use the remaining available gauge freedom (apply a suitable diffeomorphism $f(\lambda)$) in such a way 
that $K^a(\lambda)$ takes on one of the forms (s0)--(s3), (n1), (n2), with coefficients $h_i(\lambda)$.
From $\pounds_{K(\lambda)}g_{ab}(\lambda)=0$ it then follows that $\pounds_{K}\gamma_{ab}=\pounds_{\delta K}g_{ab}$, and it follows that 
at $\eH^+$, $\delta K^a =(\delta \kappa/\kappa) K^a$, whereas near $\scri$, we have, e.g., in the case (s0),  
$\delta K^a = \delta h_1 T^a + \delta h_2 \psi_{12}^a + \delta h_3 \psi_{34}^a + \dots$ (table~\ref{tab:table1}). In a general gauge compatible with our gauge conditions we would have 
more generally $\pounds_{K}\gamma_{ab}=\pounds_{\delta K}g_{ab} + \pounds_{K} \pounds_{X} g_{ab}$, where $X^a$ 
must satisfy \eref{gaugerem}. Commuting the Lie-derivative operators in the last term and using that $K^a$ Lie-derives the background 
then implies that a {\em perturbation to another stationary black hole}  is one such that 
\ben
\pounds_{K}\gamma_{ab}=\pounds_{Y}g_{ab}
\een
where $Y^a$ is such that near $\scri$, it is a sum of the form $\delta h_1 T^a + \delta h_2 \psi_{12}^a + \delta h_3 \psi_{34}^a + \dots$ [in the case (s0)] and an asymptotic symmetry from 
the subspace $\g_K$ of the preceding lemma, and where near $\eH^+$, it is of the form $Y^a = fk^a$ with $k^a\nabla_a f= 0$. 
It can be seen that the $Y^a$ in this class can realize an arbitrary asymptotic symmetry in $\so(2,d-1)$ near $\scri$. 
With our notion of perturbation towards another stationary black hole at hand (which differs slightly from the 
asymptotically flat case), we are now in a position to state:

\begin{lemma}\label{nondeglemma}
Let $\Sigma$ be a Cauchy surface for the domain of outer communications. Among the perturbations 
satisfying our gauge conditions,  together with $\delta A=0$ and $\delta H_X=0$ for all asymptotic 
symmetries $X^a$, $\E_K(\gamma, \Sigma)$ is degenerate precisely for perturbations towards other stationary black holes.
\end{lemma}
 
The proof of this lemma follows straightforwardly from lemma 2
of~\cite{Hollands:2012sf}. \qed

With the monotonicity (lemma~\ref{fluxlemma}) and non-degeneracy
(lemma~\ref{nondeglemma}) property in place, we can now make the
following argument for instability of the
background~\cite{Hollands:2012sf}.  Suppose that, on a Cauchy surface
$\Sigma$ as in \fref{DOC2}, we have $\E_K(\gamma) < 0$ for some
perturbation satisfying \eref{horgauge} as well as $\delta H_X = 0$
for all asymptotic symmetries $X \in \so(2,d-1)$. (The conditions
$\delta H_X = 0$ are of course trivially satisfied if $\gamma_{ab}$
has compact support on $\Sigma$, as will be the case in our
application below.) Then, due to the monotonicity, $\overline
\E_K(\gamma)$ cannot go to zero on any later slice $\Sigma'$ as
depicted in \fref{DOC2}. By the non-degeneracy property, it can
therefore not converge, in any sense, to a perturbation towards
another black hole background of the type considered.  Thus, the
background must be unstable.

\medskip

Later, it will be convenient to write the canonical energy in terms of the linearized initial data\footnote[2]{As usual, 
$\sqrt{q}$ is defined relative to a rigidly fixed background $(d-1)$-form e.g., defined by $d^{d-1} x$ relative to a fixed 
coordinate system on $\Sigma$. The expression for $\delta p^{ab}$ holds only for the transverse-traceless, temporal 
gauge considered in the following subsections. In a general gauge, there would be additional terms.} 
\ben 
\delta q_{ab} = q_a^{\phantom{a}c} q_b^{\phantom{b}d} \gamma_{cd}, \qquad 
\delta p^{ab} = \sqrt{q} (q^{ac} q^{bd} - q^{ab} q^{cd}) {\frac{1}{2}} \pounds_n \gamma_{cd} 
\een 
of the perturbation $\gamma_{ab}$, where $q_{ab} = g_{ab} + n_a n_b$ projects onto the tangent space of $\Sigma$. These have to satisfy the linearized constraint equations. Denoting by 
\ben
q_{ab} = g_{ab} + n_a n_b, \qquad p^{ab} = \sqrt{q} (k^{ab} -  q^{ab} k_{\phantom{c}c}^c)
\een
the initial data of the background metric $g_{ab}$ on $\Sigma$ (with $q_{ab}$ the induced metric and $k_{ab}$ the extrinsic curvature) 
and using the background constraint equations, 
the linearized constraints become, 
\begin{equation}
  \label{eq:constraints-lin}
  \fl \boldsymbol{C}(\delta q_{ab},\delta p^{ab}) \equiv 
  \left(
\begin{array}{c}
             q^{\frac{1}{2}}\left(D^aD_a\delta q_c^{\phantom{c}c} - D^aD^b\delta q_{ab} +Ric(q)^{ab}\delta q_{ab}\right) + \\
             q^{-\frac{1}{2}}\left(-\delta q_c^{\phantom{c}c} p^{ab}p_{ab} + 2p_{ab}\delta p^{ab} + 2 p^{ac}p^b_{\phantom{b}a}\delta q_{bc} + \right. \\
             \left. \frac{1}{d-2}p^c_{\phantom{c}c}p^d_{\phantom{d}d}\delta q^a_{\phantom{a}a} - \frac{2}{d-2}p^a_{\phantom{a}a}\delta p^b_{\phantom{b}b} - \frac{2}{d-2}\delta q_{ab}p^{ab}p_c^{\phantom{c}c}\right)
          \\
          \\
           -2q^{\frac{1}{2}}D^b(q^{-\frac{1}{2}}\delta p_{ab}) + D_a\delta q_{cb}p^{cb} - 2D_c\delta q_{ab}p^{bc}
\end{array} 
         \right)
       =0 \, ,
\end{equation}
where $D_a$ is the covariant derivative of $q_{ab}$.
The canonical energy can be written in terms of the background initial data $(q_{ab}, p^{ab})$, the initial data of the perturbation 
$(\delta q_{ab}, \delta p^{ab})$ and the lapse and shift $(N,N^a)$ of the Killing field, 
\ben\label{lapseshift}
K^a=Nn^a + N^a \ , 
\een
see eq.~(86) of~\cite{Hollands:2012sf}. That expression is very complicated, but it simplifies 
drastically if we assume that $(\delta q_{ab}, \delta p^{ab})$ have support in a compact subset $U \subset \Sigma$, and that $\Sigma$ is chosen so that $N=0$ 
on $U$. This is the case of interest for us, since if $K^a$ is spacelike at some point $x \in \M$,
we can choose a Cauchy surface $\Sigma$ passing through $x$ such that $K^a$ is tangent to 
$\Sigma$ in a neighborhood, $U$, of $x$; as we shall see, we can then construct WKB initial
data with support in $U$. The resulting expression is:
\begin{eqnarray}
  \label{eq:canonical-energy}
  \fl
 \mathcal{E}_K(\delta q_{ab},\delta p^{ab}) =  -\frac{1}{16\pi}\int_\Sigma N^a\left( -2 \delta p^{bc} D_a \delta q_{bc} + 4 \delta p^{cb} D_b\delta q_{ac} +2 \delta q_{ac}D_b\delta p^{cb} \right. \nonumber\\
  \left.- 2 p^{cb}\delta q_{ad} D_b\delta q_c^{\phantom{c}d} + p^{cb} \delta q_{ad} D^d \delta q_{cb} \right) .
\end{eqnarray}

\section{High frequency gravitational waves}\label{wkb}

In order to construct a perturbation with $\E_K(\gamma, \Sigma) <0$ 
on our AdS black hole background, we use a high frequency (WKB) ansatz for the gravitational perturbation $\gamma_{ab}$
(for further discussion, see,
e.g.,~\cite{Isaacson:1967zz}, or section III.12
of~\cite{Choquet-Bruhat:2009}). The ansatz is, as usual, 
\ben\label{WKB}
\eqalign{\gamma_{ab}(x,\omega) = A_{ab}(x,\omega) \exp(i\omega \chi(x)), 
\\ A_{ab}^{}(x,\omega) = \sum_{n \ge 0} A^{(n)}_{ab}(x) (i\omega)^{-n}_{}, }
\een 
where $\omega \gg 1$, where the phase function $\chi$ is a smooth real valued function on $\M$ and where the $n$-th order amplitudes $A_{ab}^{(n)}$ 
are smooth real valued symmetric tensor fields on $\M$. We can take the real part of $\gamma_{ab}$ in order to obtain a real-valued perturbation in the end. The sum 
is understood in the sense of an asymptotic series in $1/\omega$; it is not expected 
to converge but that is not of concern here since
we will use only a finite number of terms to generate a suitable approximate solution.
Substituting the WKB ansatz into the linearized Einstein equation~\eref{ein} in the transverse ($\nabla^a \gamma_{ab} = 0$) and traceless ($\gamma^a_{\phantom{a}a} = 0$) 
gauge\footnote[3]{If no gauge conditions are imposed but it is assumed that $\gamma_{ab}$ is not pure gauge, then one still obtains \eref{eikonal}
together with a ``polarization condition'' on $A_{ab}$; see~\cite{Choquet-Bruhat:2009} for further
discussion.} yields the usual 
eikonal equation
\ben\label{eikonal}
g^{ab} (\nabla_b \chi)\nabla_a \chi = 0 
\een
and transport equation
\ben\label{transport0}
\left(2(\nabla^b\chi) \nabla_b + \nabla^b \nabla_b \chi\right) A_{ac}^{(0)} =0
\een
as well as the transport equations
\ben\label{transport}
\left(2(\nabla^b\chi) \nabla_b + \nabla^b \nabla_b \chi\right) A_{ac}^{(n+1)} =
 - \nabla^b \nabla_b A_{ac}^{(n)} - 2R_a{}^b{}_c{}^d A_{bd}^{(n)} 
\een
for $n \ge 0$.
The eikonal equation states that the surfaces of constant phase, $\chi$, (i.e., the wave fronts) are null, 
and it follows immediately that $p_a \equiv \nabla_a \chi$ is tangent to null geodesics, 
$p^a \nabla_a p^b = 0$. We can always find solutions to the eikonal equation  
locally\footnote[4]{As usual, 
the WKB ansatz breaks down where the congruence $p^a$ forms caustics, but this is of no 
relevance for us because we are only interested in a local solution near $\Sigma$.}
near a given point $x \in \Sigma$. For instance, we can construct a $\chi$ in a neighborhood of
$x$ by starting with 
a $(d-2)$-dimensional embedded hypersurface $S$ in $\Sigma$ passing through $x$. 
Near $S$, we can introduce Gaussian normal coordinates within $\Sigma$ in the form $q_{ab} = s_{ab}(\chi) + (D_a \chi)D_b \chi$, where $\chi$ is the coordinate 
transverse to $S$ describing a local foliation $\{S_\chi\}_{\chi \in \RR}$ of surfaces labelled by a parameter $\chi$ such that $S_0=S$, where $s_{ab}(\chi)$ defines 
a metric on each $S_\chi$. 
Consequently $q^{ab}(D_b \chi)D_a \chi =1$, meaning that the vector field  $p^a \equiv n^a + q^{ab} D_b \chi$ defined on $\Sigma$ is null there. We extend this 
to a null field off $\Sigma$ via the geodesic equation $p^b \nabla_b p^a = 0$. Moving the surfaces $S_\chi$ along these null geodesics off $\Sigma$ defines 
null surfaces, and we define $\chi$ in a neighborhood of $\Sigma$ to be constant along each such null surface. It follows that $\chi$ solves the eikonal equation 
and $\nabla_a \chi = p_a$ near $\Sigma$. In addition we have $n^a  p_a = -1$ on $\Sigma$. 

The transport equation allows the recursive determination of the tensor coefficients 
in the series for $A_{ab}$ by solving an ordinary differential equation along the orbits of $p^a$. 
Of course, we also need to satisfy the gauge conditions, which become
\ben\label{gauge}
A_a \equiv i\omega p^b A_{ab} + \nabla^b A_{ab}=0 \ , \qquad
A \equiv A^a{}_a = 0 \ . 
\een
Thus, given a solution $\chi$ to the eikonal equation, we wish to find a solution $A_{ab}$
to the transport equations~\eref{transport},~\eref{transport0} near $\Sigma$ (in the sense of an asymptotic series in $1/\omega$) such that the gauge conditions \eref{gauge} hold order-by-order.  

We do this as follows. First, we note that the eikonal and transport equations for $\chi$ and $A_{ab}^{(n)}$, the 
background Einstein equations, and the Bianchi identity imply the transport equations
(remembering $p_a = \nabla_a \chi$)
\ben\label{transport1}
\eqalign{
(2p^b \nabla_b + \nabla^b p_b) A_{a}^{(n+1)} &=  -\nabla^b \nabla_b A_{a}^{(n)} - \frac{2}{d-2} \Lambda A_{a}^{(n)}   , \cr
(2p^b \nabla_b + \nabla^b p_b)  A^{(n+1)}  &=  -\nabla^b \nabla_b A^{(n)} - \frac{4}{d-2} \Lambda A^{(n)}  , 
}
\een
for the expansion coefficients of the ``gauge conditions'' $A_a$ and $A$\footnote[5]{In other words, 
$A^{(0)}_a \equiv p^a A^{(0)}_{ab}$ and $A^{(n+1)}_a \equiv p^a A^{(n+1)}_{ab}+\nabla^b A^{(n)}_{ab}$ for $n \ge 0$, as well as $A^{(n)} \equiv A^{(n)}{}^a_{\phantom{a}a}$.}
for $n \ge -1$. (For $n=-1$, the right side is by definition equal to zero.)  
Thus, we can recursively (in the WKB expansion) 
satisfy the gauge conditions~\eref{gauge} if we satisfy them on $\Sigma$. Indeed, by choosing $A^{(0)}_{ab}$ to be any
 symmetric tensor defined on $\Sigma$ such that 
\ben\label{A0}
A^{(0)}_{ab} p^a = 0 \ , \qquad A^{(0)}_{ab} g^{ab} = 0 \ , \qquad A^{(0)}_{ab} n^b = 0 
\ , \qquad \mathrm{on}\ \Sigma,
\een 
we clearly satisfy all the gauge conditions on $\Sigma$ (and additionally the ``temporal gauge'' expressed by the last equation)
at zeroth WKB order. We ``evolve'' this $A_{ab}^{(0)}$ off of $\Sigma$ using the 
transport equation~\eref{transport0}. Then the leading order $n=-1$ transport equations~\eref{transport1} imply that the above gauge conditions~\eref{A0} 
(except the temporal gauge) hold in a neighborhood of $\Sigma$. Inductively, we choose at $(n+1)$-th WKB order any $A^{(n+1)}_{ab},n \ge 0$ 
such that 
\ben\label{An}
A^{(n+1)}_{ab} p^b + \nabla^b A^{(n)}_{ab} = 0 \ , \quad
A^{(n+1)}_{ab} g^{ab} = 0 \ , \quad A^{(n+1)}_{ab} n^b = 0 \ ,  \quad \mathrm{on}\ \Sigma,
\een
and we extend $A^{(n+1)}_{ab}$ off of $\Sigma$ with the $n$-th order transport equation~\eref{transport}. The $n$-th order transport equations~\eref{transport1}
then imply that the $(n+1)$-th order gauge conditions~\eref{An} (except the temporal gauge condition) also hold in a neighborhood of $\Sigma$. Note that the 
conditions~\eref{A0} and~\eref{An} are entirely algebraic at each order. In particular, we can arrange all expansion coefficients of $A_{ab}$ to have support 
in a ``light like tube'' obtained by moving some arbitrarily chosen compact subset  $U\subset \Sigma$ along the orbits of $p^a$. Below, we will choose $U$ to 
be a small neighborhood of a point $x \in \M$ where $K^a$ is space like. 

We can also describe the WKB approximate solutions in terms of their initial data on $\Sigma$.  For a solution $\gamma_{ab}$ of WKB form~\eref{WKB} 
just described, the initial data have the expansions
\ben\label{pqdef}\eqalign{
  \delta q_{ab} = \left( \sum_{n \ge 0} Q^{(n)}_{ab} (i\omega)^{-n}_{} \right) \exp(i\omega \chi) , 
  \\
  \delta p_{ab} = \left( \sum_{n \ge 0} P^{(n)}_{ab} (i\omega)^{-n+1}_{} \right) \exp(i\omega \chi) . }
\een
Our ansatz~\eref{WKB}, the leading order gauge conditions~\eref{gauge}, and the condition $n^a \nabla_a \chi = -1$ on $\Sigma$ 
imply that 
\ben\label{wkbzero}
P^{(0)}_{ab} = -Q^{(0)}_{ab} , \qquad
Q^{(0)a}_a=0,\qquad 
Q^{(0)}_{ab}D^b\chi  = 0 ,  
\een
and with this, the leading order linearized constraints~\eref{eq:constraints-lin} (order $\omega^2$) are satisfied, as they must be. 
The higher order constraints must also be satisfied for initial data coming from a WKB perturbation of the form~\eref{WKB}
as described. [Alternatively, we could use the linearized constraints at higher WKB orders directly to derive the algebraic conditions on 
the higher order tensors $(Q_{ab}^{(n)}, P^{(n)}_{ab})$: At $n$-th order, the linearized constraints take the form: 
\begin{equation}
  \label{eq:WKB-n}
  \left( \begin{array}{c}
           -D^a\chi (D_a\chi)Q^{(n)c}_c + D^a\chi (D^b\chi) Q^{(n)}_{ab} \\
           P^{(n)}_{ab} D^b\chi
         \end{array} \right) = \boldsymbol{C}^{(n)},
\end{equation}
where the source $\boldsymbol{C}^{(n)}$ depends on the lower order WKB approximations
$(Q_{ab}^{(m)},P^{(m)}_{ab})$ for $m<n$. As before, the  left side is algebraic in the fields, so it is possible to 
maintain support within an arbitrary compact subset $U \subset \Sigma$.]

\section{Proof of theorem \ref{thm1}} \label{neg}

We now have all the ingredients for the proof of
theorem~\ref{thm1}. Let $x$ be a point in the ergoregion, i.e.,
$K^a|_x$ is spacelike. We can then pick a Cauchy surface $\Sigma$ such
that $K^a$ is tangent to $\Sigma$ within some sufficiently small open
subset $U \subset \Sigma$ containing $x$. Consequently, in that
subset, the lapse $N=0$, see eq.~\eref{lapseshift}. Let $(\delta
q_{ab},\delta p^{ab})$ be initial data on $\Sigma$ given by the real part of
the WKB form~\eref{pqdef}, with the WKB expansion carried out up to
some finite order $n$; the value of $n$ will be chosen later.  We can
arrange the initial data to be smooth and have compact support in $U$,
and the zeroth order WKB expansion coefficients to satisfy the
algebraic constraints~\eref{wkbzero}, and---as we explained---we can
also choose $\chi$ such that $n^a \nabla_a \chi = -1$ on $U$.  Since
the lapse $N=0$ in $U$, the canonical energy $\E \equiv \E_K(\gamma,
\Sigma)$ can be written as in eq.~\eref{eq:canonical-energy}, which to
leading order in the WKB parameter $\omega$
gives\footnote[6]{\label{footnote6} Note that if $Q^{(0)}_{ab}$ is
  chosen to be sharply peaked near $x$, the right side is
  approximately $-({\rm cst.}) \omega^2 K^a p_a|_x$ where cst. $\sim
  \frac{1}{16\pi} Q^{(0)a}_c Q^{(0)c}_a |_x$ is a positive constant.}
(remembering $p_a = \nabla_a \chi$, and using that $\sin^2 (\omega
\chi) \to \frac{1}{2}$ weakly as $\omega \to \infty$)
\ben\label{Eineq} \mathcal{E}(\delta q, \delta p)
  = -\frac{\omega^2}{16\pi} \int_U K^b p_b \, Q^{(0)a}_c Q^{(0)c}_a  +O(\omega)  \ . 
\een
The explicitly written $O(\omega^2)$-term\footnote[7]{
Our use of the ``big-O'' notation here is the following. We write $f(\omega) = O(\omega^{-k})$ if it is true that 
$\lim_{\omega \to \infty} \omega^{k-\delta} |f(\omega)| = 0$ for each $\delta > 0$. If $f(x,\omega)$ is also a (smooth) function 
or tensor field of $x \in \Sigma$, we write $f(\omega,x) = O(\omega^{-k})$ if it is true that 
$\lim_{\omega \to \infty} \omega^{k-\delta-j} |D^j f(x, \omega)|_q = 0$ for each $\delta > 0$ and each $j=0,1,2, \dots$.
} dominates for $\omega \gg 1$ and 
is manifestly negative provided $K^a p_a > 0$ in $U$, which we can always arrange by 
a suitable choice of $U$ and $\chi$. Thus, we have constructed compactly supported initial data such that $\E<0$. 

However, we are not done with the proof yet, because these initial data only correspond to an approximate solution in the WKB sense, and not an exact one. 
In other words, the linearized constraints  \eref{eq:constraints-lin} are not satisfied but instead we have 
\ben
\boldsymbol{C}(\delta q, \delta p) = \boldsymbol{J}  = \left( \sum_{k=n-1}^{n} (i\omega)^{-k} \boldsymbol{J}^{(k)} \right) \exp(i\omega \chi) \ , 
\een
where each $\boldsymbol{J}^{(k)} = (u^{(k)}_{\phantom{(m)}}, X^{(k)}_a)$ is a pair of a scalar density and a dual vector density on $\Sigma$ 
that is constructed out of the WKB expansion tensors $(Q_{ab}^{(m)},P^{(m)}_{ab})$ for $m\le n$. In particular, each such tensor 
has compact support in $U \subset \Sigma$. We wish to correct our WKB initial data $(\delta q_{ab}, \delta p^{ab})$ in such a way that 
\begin{enumerate}
\item The linearized constraints hold exactly. 
\item The data remain smooth and compactly supported in a somewhat larger open region $V$ containing the closure of $U$.
\item The correction has a $H^k$-Sobolev norm of order $O(\omega^{-n+1+k})$ as $\omega \to \infty$. 
\end{enumerate}
The first two items imply that the corrected initial data can be used to make our instability argument, and the third implies that the canonical energy 
of the corrected initial data is still negative for sufficiently large $\omega \gg 1$ provided the WKB order $n$ is chosen to be sufficiently large, 
because the canonical energy is a continuous quadratic form on 
the Sobolev space $H^1$ (it depends on at most one derivative of the linearized initial data on $V$). We now explain the details. 

Following~\cite{Corvino:2003sp,Chrusciel:2003sr}, the idea is to make a particular ansatz for the 
correction to $(\delta q_{ab}, \delta p^{ab})$. The linearized constraints $\boldsymbol{C}$ may be viewed as the
result of acting on the perturbed initial data by a linear operator which 
maps the pair $(\delta q_{ab}, \delta p^{ab})$ consisting of a symmetric tensor, $\delta q_{ab}$,
and a symmetric tensor density, $\delta p^{ab}$, on $\Sigma$ into a pair $(u,X_a)$ consisting of a scalar density
and dual vector density on $\Sigma$. Therefore, its adjoint
differential operator, $\boldsymbol{C}^*$, maps a pair $\bX=(u,X^a)$ consisting of
a scalar and vector field on $\Sigma$ into a pair $(\delta q_{ab}, \delta p^{ab})$ consisting of a symmetric tensor density and symmetric tensor on $\Sigma$.
One can straightforwardly calculate that $\boldsymbol{C}^*$ is given by
\ben\label{delc*}
 \boldsymbol{C}^*  \left(
\begin{array}{c}
u\\
X^a
\end{array}
\right) = \left(
\begin{array}{c}
q^\half(-(D^c D_c u) q^{ab} + D^a D^b u  + Ric(q)_{ab} u)+\\
q^{-\half}( - q^{ab} p^{cd} p_{cd} u + 2 p^{(a}{}_c p^{b)c} u + \frac{1}{d-2} q^{ab} p^c{}_c p^d{}_d u\\
-\frac{2}{d-2}p^{ab}p^c{}_c u - p^{ab} D_c X^c + 2D_c X^{(a} p^{b)c})\\
\\
q^{-\half}(2p_{ab}u
- \frac{2}{d-2} q_{ab} p^c{}_c u)
+\pounds_X q_{ab}
\end{array}
\right).
\een
Let $s: V \to \RR$ be a function $1 \ge s>0$ such that near the boundary $\partial V$, we have
\ben
s(x) = {\rm dist}_q(x, \partial V) \ ,
\een
where we mean the geodesic distance relative to the Riemannian metric $q_{ab}$ on $\Sigma$. We also ask that $s(x) = 1$ in $U \subset V$.
The ansatz for the corrected linearized initial data is:
\ben\label{ansatz}
\left(
\begin{array}{c}
\delta \tilde q_{ab}\\
\delta \tilde p^{ab}
\end{array}
\right) \equiv
\left(
\begin{array}{c}
\delta q_{ab}\\
\delta p^{ab}
\end{array}
\right)
-
e^{-2/s^\alpha} \left(
\begin{array}{cc}
s^{4\alpha+4} & 0\\
0 & s^{2\alpha+2}
\end{array}
\right)
 \boldsymbol{C}^* \left(
\begin{array}{c}
u \\
X^a
\end{array}
\right)
\een
where $\alpha>0$ is later chosen to be sufficiently large.
Cutoff functions involving $s$ have been inserted because we hope to extend
the solution by $0$ across the boundary $\partial V$ in a smooth way. The tensors $\bX \equiv (u, X^a)$ are to be determined.
 For the matrix of
 cutoff functions  we introduce the shorthand:
\ben\label{matrixm}
\Phi \equiv e^{-1/s^\alpha} \left(
\begin{array}{cc}
s^{2\alpha+2} & 0\\
0 & s^{\alpha+1}
\end{array}
\right).
\een
Our ansatz can then be written in a more condensed fashion as
\ben\label{corrected}
\left(
\begin{array}{c}
\delta \tilde q\\
\delta \tilde p
\end{array}
\right) =
\left(
\begin{array}{c}
\delta q\\
\delta p
\end{array}
\right)
-
\Phi^2 \boldsymbol{C}^* \bX \ .
\een
We want $(\delta \tilde q_{ab}, \delta \tilde p^{ab})$ to satisfy the
linearized constraints. Acting with $\boldsymbol{C}$ shows that
$\bX$ must satisfy the fourth order mixed elliptic system of equations:
\ben\label{xeps}
\boldsymbol{C} \Phi^2 \boldsymbol{C}^* \bX  = \boldsymbol{J} \ .
\een
It was shown in lemma~6.2 of~\cite{Hollands:2014lra} that 
there exists a smooth solution $\bX$ to~\eref{xeps} in $V$ which
additionally satisfies for all $k=0,1,2,\dots$
\ben\label{regularity}
\int_V s^{2k\beta} |D^k (\Phi \boldsymbol{C}^* \bX)|^2_q  \le c_1 \| \boldsymbol{J} \|_{H^k}^2
\een
for suitably large $\beta>\alpha>0$, and a constant $c_1=c_1(V,\alpha,\beta, k)$. Due to the exponential factor in $\Phi$, it follows 
 in particular that $\Phi^2 \boldsymbol{C}^* \bX$ (note the {\em square} in $\Phi^2$ compared to~\eref{regularity}) is smooth up to and including the boundary $\partial V$, and that it can in fact be
smoothly extended by $0$ across $\partial V$, see remark a) following lemma~6.2 of~\cite{Hollands:2014lra}. 
Thus, the corrected initial data~\eref{corrected} are smooth up to and
including the boundary $\partial V$ and can be smoothly extended by $0$ across $\partial V$. As a consequence of~\eref{regularity}, 
we also have
\ben\label{estimate}
\| \delta p - \delta \tilde p\|_{H^k} + \|\delta q - \delta \tilde q\|_{H^k} \le c_2 \| \boldsymbol{J}\|_{H^k} = O(\omega^{-n+1+k}) \  
\een
for some constant $c_2$. By definition we have $\|\delta p\|_{H^k} = O(\omega^{k+1}), \|\delta q\|_{H^k} = O(\omega^k)$, since 
$\delta p^{ab}$ is of order $O(\omega)$, $\delta q_{ab}$ is of order $O(\omega^0)$, and each derivative brings down one factor of $\omega$, i.e., 
in total the factor $\omega^k$ from the $k$ derivatives in the $H^k$ norm. 
Since $\E$ is a quadratic form depending on up to one derivative, it follows via the Cauchy-Schwarz inequality that
\begin{eqnarray}
 \E(\delta \tilde q, \delta \tilde p) &\le & 
\E(\delta q, \delta p) \nonumber\\ 
&&+ c_3 (\|\delta p\|_{H^1} + \| \delta q \|_{H^1})(\| \delta p - \delta \tilde p\|_{H^1} + \|\delta q - \delta \tilde q\|_{H^1}  ) \nonumber\\
&&+ c_4 (\| \delta p - \delta \tilde p\|_{H^1} + \|\delta q - \delta \tilde q\|_{H^1}  )^2\nonumber\\
&\le& \E(\delta q, \delta p) + O(\omega)
\end{eqnarray}
where in the last line we have chosen $n \ge 3$. Combining this inequality with~\eref{Eineq}, we see 
that $\E(\delta \tilde q, \delta \tilde p) <0$ for sufficiently large $\omega$. The proof is complete. 

\section{Discussion}\label{sec:discussion}

We have shown that any asymptotically AdS black hole with an
ergoregion is linearly unstable. This immediately implies that
Kerr-AdS is unstable to gravitational perturbations for rotation
speeds above the Hawking-Reall bound~\cite{Hawking:1999dp}. The
recently discovered ``black resonator'' solutions~\cite{Dias:2015rxy,Niehoff:2015oga}, which have a single
helical Killing field and always contain an ergoregion, are also
unstable. 

We have restricted consideration
in this paper to vacuum spacetimes. However, 
since the essential properties of canonical energy needed for our analysis
follow directly from the Lagrangian
formulation as well as positivity of flux through the horizon, it should be straightforward to 
extend our analysis to show a similar ergoregion instability when matter fields are present, in
particular electromagnetic~\cite{Hollands:2014lra} and scalar
fields~\cite{Keir:2013jga}.

Another possible generalization of our work concerns the case of charged black
holes. For a test particle of mass $m$ and charge $q$ in the spacetime
of a charged black hole, the $4$-momentum of the
particle is given by
\ben
p_a = m u_a + q A_a
\een
where $u^a$ is the $4$-velocity of the particle and $A_a$ is
the vector potential of the black hole. For such a charged particle, the region of spacetime
where the energy
\ben
\E_{K, \rm particle} \equiv -K^a p_a = - m K^a u_a - q K^a A_a \ ,
\een
may be made negative is called the ``generalized
ergoregion''~\cite{Christodoulou:1972kt,Denardo:1973}. Generalized ergoregions
can occur even in cases where $K^a$ is everywhere timelike in the exterior region, 
such as for a Reissner-Nordstr\"om-AdS black hole.
If the energy of a charged {\em field} can also be made negative for a black hole
with a generalized ergoregion, then a superradiance
phenomenon similar to the rotating case can occur.
For a charged black hole in an asymptotically AdS spacetime\footnote[8]{In the asymptotically flat case,
in order to get a positive flux one must impose a gauge condition that cannot be
simultaneously satisfied at both $\mathscr{H}^+$
and $\scri^+$, so, as for rotating black holes, superradiance does not imply instability in this case.}, this would give rise to an 
instability~\cite{Gubser:2008px,Basu:2010uz}.

One might expect that a charged black hole in an asymptotically AdS spacetime would 
be unstable to perturbations of a field of mass $m$ and charge $q$ whenever a generalized
ergoregion exists for particles of the same mass and charge. However, this is not the case, since,
unlike the rotating case, the existence of a generalized ergoregion for particles does not
imply that initial data for a field of the same mass and charge 
parameters can be chosen to have negative energy. As we have shown in this
paper, high frequency
gravitational wave initial data can be constructed that has properties arbitrarily close to that
of a null particle. Thus, initial data with negative canonical energy can be constructed whenever 
$K^a$ is spacelike, i.e., whenever there exists a particle ergoregion. However, if one performs
a WKB analysis to construct initial data for a charged scalar field analogous to that of \sref{wkb}, 
one does not obtain useful results. This is because the charge of a scalar field
is given by an integral over its charge-current vector, involving one spacetime
derivative, whereas its energy involves two derivatives. Consequently, in the high-frequency
limit, the charge to mass ratio of the wavepacket 
goes to zero, and one cannot take advantage of the 
negative electromagnetic contribution to the total energy. Thus, it is not useful to make a high frequency
approximation when searching for initial data for charged fields with negative energy.
In fact, for a Reissner-Nordstr\"om-AdS black hole, the instability for a scalar field of mass
$m$ and charge $q$ sets in for slightly different black hole parameters than
the appearance of a generalized ergoregion for point particles of the same mass and charge. We
have checked numerically that the onset of instability through the
canonical energy method is the same as that identified
in~\cite{Gubser:2008px}.

\ack

The work of S.H. is supported in part by ERC starting grant
259562. The work of A.I. was supported in part by JSPS KAKENHI Grants
No.~15K05092 and No.~23740200. This research was also supported in
part by NSF grants PHY-1202718 and PHY-1505124 to the University of
Chicago and by Perimeter Institute for Theoretical Physics. Research
at Perimeter Institute is supported by the Government of Canada
through Industry Canada and by the Province of Ontario through the
Ministry of Research and Innovation.

\appendix

\section{Orbits in $\frak{so}(2,d-1)$ and normal forms}\label{appendix}

We realize $\so(2,d-1)$ by real matrices $X$ of size $d+1$ with the property that 
${}^t X\eta + \eta X = 0$, where $\eta = diag(-1,-1,1, \dots, 1)$. An element $X$ of 
a real Lie algebra $\g$ is called semi-simple if $\ad X = [X, \ . \ ]$ is diagonalizable (in $\g_\CC$). 
$X$ is called nil-potent if $\ad X$ is nil-potent. It follows from Chevalley's theorem 
that any $X \in \so(2,d-1)$ has a unique decomposition $X=X_s + X_n$ into a semi-simple 
and a nil-potent part, both of which lie in $\so(2,d-1)$. The adjoint action of $G={\rm SO}_+(2,d-1)$ 
on $\so(2,d-1)$ is denoted by $g \cdot X = {\rm Ad}(g) X$. A {\em Cartan subalgebra}, $\h$ 
is a real maximally abelian sub algebra such that any element $X \in \h$ is semi-simple. Two Cartan 
subalgefbras $\h_1, \h_2$ are called {\em conjugate} if there is a $g \in G$ such that $g \cdot \h_1 = \h_2$. 
Two Cartan subalgebras are called {\em inequivalent} if they are not conjugate to each other. Let $N$ 
be the number of inequivalent Cartan subalgebras and denote by $\h_1, \dots, \h_N$ canonical representatives, 
i.e., any other Cartan sub algebra is conjugate to exactly one of these. It is known (see Para.~3 of~\cite{sig}, see also~\cite{kost}) 
that for $\so(2,d-1)$, 
$N=3$ for odd $d$ and $N=4$ for even $d$.  

We are interested in classifying the $G$-orbits in $\so(2,d-1)$. We first consider {\em regular} orbits $G \cdot X$, i.e., 
ones with the maximum possible dimension (such $X$ are called regular, too). Our aim is to identify for each such orbit a
canonical representative, which we think of as a normal form. 

{\bf Case 1)} Assume that $X=X_s$, i.e., that $X$ is semi-simple. It follows from \cite{roth1} that 
$X$ is conjugate to an element $H = g \cdot X$ in precisely one of the canonical Cartan sub algebras $\h_1, \dots, \h_N$ of $\so(2,d-1)$
displayed explicitly in Para.~3 of~\cite{sig}. Based on this classification, one arrives at the following 
canonical representatives for $X$: 
\begin{enumerate}

\item[s0)] $X$ is conjugate under $G$ to an $H$ of the form
\ben
H=
\left(
  \begin{array}{ccccccccc}
    0 & h_1 &      &     &      &     & & &\\
  h_1 & 0   &      &     &      &     & & &\\
      &     & 0    & h_2 &      &     & & &\\
      &     & -h_2 & 0   &      &     & & &\\
      &     &      &     & 0    & h_3 & & &\\
      &     &      &     & -h_3 & 0   &        &     & \\
      &     &      &     &      &     & \ddots &     &  \\
      &     &      &     &      &     &        & 0   & h_m\\
      &     &      &     &      &     &        &-h_m & 0 
  \end{array}
\right)
 \ , 
\een
where $h_i \in \RR, h_i \neq 0$ are mutually distinct and $m=\lfloor (d-1)/2 \rfloor$. When $d$ is even 
there is one additional last row and column of zeros.

\item[s1)] $X$ is conjugate under $G$ to an $H$ of the form
\ben
 H=
\left(
\begin{array}{ccccccccc}
0 & -h_2 & h_1 & 0 &  &  & & & \\
h_2 & 0 & 0 &  h_1 &  &  &  & & \\
h_1 & 0 & 0 & h_2 &  &  &  & &\\
0 & h_1 & -h_2 & 0 &  &  &   & &\\
 &      &       &  & 0 & h_3&  & &\\
 &      &       &  & -h_3 & 0&  & & \\
 & & & & & & \ddots & &  \\
& & & & & & & 0  & h_m\\
& & & & & & & -h_m & 0 
\end{array}
\right)
\ , 
\een
where $h_i \in \RR, h_i \neq 0$ are mutually distinct and $m=\lfloor (d-3)/2 \rfloor$. When $d$ is even 
there is one additional last row and column of zeros. 

\item[s2)] $X$ is conjugate under $G$ to an $H$ of the form
\ben
 H=
\left(
\begin{array}{ccccccccc}
0 & 0 & h_1 & 0 &  &  & & & \\
0 & 0 & 0 &  h_2 &  &  &  & & \\
h_1 & 0 & 0 & 0 &  &  &  & &\\
0 & h_2 & 0 & 0 &  &  &   & &\\
 &      &       &  & 0 & h_3&  & &\\
 &      &       &  & -h_3 & 0&  & & \\
 & & & & & & \ddots & &  \\
& & & & & & & 0  & h_m\\
& & & & & & & -h_m & 0 
\end{array}
\right)
\ , 
\een
where $h_i \in \RR, h_i \neq 0$ are mutually distinct and $m=\lfloor (d-3)/2 \rfloor$. When $d$ is even 
there is one additional last row and column of zeros. 

\item[s3)] $X$ is conjugate under $G$ to an $H$ of the form
\ben
H=
\left(
\begin{array}{ccccccccc}
0 & 0 &  0 &  &  & & & \\
0 & 0 &  h_1 &  &  &  & & \\
0 & h_1 & 0 &  &  &  & &\\
 &      &         & 0 & h_2&  & &\\
 &      &         & -h_2 & 0&  & & \\
 & & & & &  \ddots & &  \\
& & & & & &  0  & h_m\\
& & & & & &  -h_m & 0 
\end{array}
\right)
 \ , 
\een
where $h_i \in \RR, h_i \neq 0$ are mutually distinct. This case only exists when $d$ is even and $m=(d-2)/2$. \end{enumerate}

For non-regular semi-simple $X$, there is a representer taking one of the canonical forms s0)-s3) with no restriction on the $h_i$. 

\medskip
\noindent
{\bf Case 2)} Assume that $X=X_s+X_n$ with non-zero $X_n$, i.e., that $X$ is not semi-simple. It is shown in Prop.~5.1 of~\cite{roth2} that 
$X$ is regular if and only if $\g=\so(2,d-1)^{X_s}=\{Z \mid [Z, X_s]=0\}$ contains a regular semi-simple element $Y$ such that $\ad Y$ only has real eigenvalues. 
Since $X_s$ must be in one of the Cartan subalgebras given in Case 1) (up to conjugation), we may analyze the cases in which such a $Y$ 
exists and determine the possible $X_n$. There are two cases:
\begin{enumerate}
\item[n1)] $X$ is conjugate under $G$ to $H$ given by 
\ben
H=
\left(
\begin{array}{ccccccccc}
0 & h_1 &  h_1 &  &  & & & \\
-h_1 & 0 &  0 &  &  &  & & \\
h_1 & 0 & 0 &  &  &  & &\\
 &      &         & 0 & h_2&  & &\\
 &      &         & -h_2 & 0&  & & \\
 & & & & &  \ddots & &  \\
& & & & & &  0  & h_m\\
& & & & & &  -h_m & 0 
\end{array}
\right)
 \ , 
\een
where $h_i \in \RR, h_i \neq 0$ are mutually distinct and $m=\lfloor (d-2)/2 \rfloor$. When $d$ is 
odd there is one additional last row and column filled by zeros.

\item[n2)] $X$ is conjugate under $G$ to $H$ given by 
\ben
H=
\left(
\begin{array}{ccccccccc}
0 & h_1 & h_1 & 0 &  &  & & & \\
-h_1 & 0 & 0 &  h_2 &  &  &  & & \\
h_1 & 0 & 0 & 0 &  &  &  & &\\
0 & h_2 & 0 & 0 &  &  &   & &\\
 &      &       &  & 0 & h_3&  & &\\
 &      &       &  & -h_3 & 0&  & & \\
 & & & & & & \ddots & &  \\
& & & & & & & 0  & h_m\\
& & & & & & & -h_m & 0 
\end{array}
\right)
\ , 
\een
where $h_i \in \RR, h_i \neq 0$ are mutually distinct and $m=\lfloor (d-3)/2 \rfloor$. When $d$ is even 
there is one additional last row and column of zeros. 
\end{enumerate}

For non-regular non semi-simple $X$, there is a representer taking one of the canonical forms n1),n2) with no restriction on the $h_i$. 

\medskip

The elements $X \in \so(2,d-1)$ are in 1-to-1 correspondence with asymptotic symmetries in asymptotically AdS spacetimes of dimension $d$. 
That correspondence is most easily explained in the case of exact AdS,  presented as the universal cover of the ``hyperboloid'' 
$x_0^2 + x_d^2 -x_1^2 - \dots -x_{d-1}^2   = \ell^2$ in $\RR^{d-1,2}$. In those coordinates, the matrix $X=(X^A{}_B) \in \so(2,d-1)$ with $A,B=0,d,1\dots,d-1$
corresponds to the Killing field
\ben
X = \sum_{A,B} X^A{}_B \ x_A \frac{\partial}{\partial x_B} . 
\een
A basis is, with $1\le i<j\le d-1$:
\ben\label{generators}
\eqalign{
T &= x_0 \frac{\partial}{\partial x_d} - x_d \frac{\partial}{\partial x_0},\\
C_i &= x_d \frac{\partial}{\partial x_i} + x_i \frac{\partial}{\partial x_d},\\ 
P_i &= x_0 \frac{\partial}{\partial x_A} + x_i \frac{\partial}{\partial x_0}, \\
\psi_{ij} &= x_i \frac{\partial}{\partial x_j} - x_j \frac{\partial}{\partial x_i}.
}
\een
These formulae remain true in asymptotically AdS spacetimes if we cover the asymptotic region with the same type of coordinates as pure AdS. 
The normal forms for $X$ given in s0)-s3) and n1),n2) lead to the following lemma:

\begin{lemma}\label{normformlemma}
Let $X^a$ be an (infinitesimal) asymptotic symmetry. Then there exists a diffeomorphism $f$ of $\tilde \M$ which is an asymptotic symmetry such that 
$f_* X^a$ takes one of the following forms, where $h_i \in \RR$:
\end{lemma}
\begin{table}[h!]
  \begin{center}
    \caption{Different normal forms for asymptotic symmetries.}
    \label{tab:table1}
    \begin{tabular}{l|c|r}
    Type & Normal form & Remark \\
    \hline \hline
        s0 & $h_1 T^a + h_2 \psi_{12}^a + h_3 \psi_{34}^a + h_4\psi^a_{56} + \dots $ & \\
      \hline
        s1 & $h_1(P_1^a + C_2^a) + h_2 (T^a + \psi_{12}^a) +  h_3 \psi_{34}^a + h_4 \psi_{56}^a + \dots$. \\
        \hline
       s2 & $h_1 P_1^a + h_2 C_2^a + h_3 \psi_{34}^a + h_4 \psi_{56}^a + \dots$ & \\
       \hline
       s3 & $h_1 C_1^a + h_2 \psi_{23}^a + h_3 \psi_{45}^a + \dots$ & only odd $d$ \\
       \hline
       n1 & $h_1(T^a + P^a_1) + h_2 \psi_{23}^a + h_3 \psi_{45}^a + \dots$ & \\
       \hline
       n2 & $h_1(T^a + P^a_1) + h_2 C_2^a + h_3 \psi_{34}^a + h_4 \psi_{56}^a + \dots$ & \\
    \end{tabular}
  \end{center}
\end{table}

{\em Proof:} The asymptotic symmetry $f$ acts on $\scri$ as a conformal transformation of $\RR \times S^{d-2}$ and is hence represented by 
an element $g_f \in \widetilde G$. Similarly, the restriction of $X^a$ to $\scri$ is a conformal Killing vector field of $\RR \times S^{d-2}$ and can be 
identified with an element  $X \in \so(2,d-1)$. The pull back $f_* X^a$ corresponds to the adjoint 
action of $g_f \cdot X$ under these identifications. Obviously, any $g \in G$ can be obtained in this way 
from a suitable $f$, so the lemma follows from our previous discussion of the $G$-orbits in $\so(2,d-1)$.

\bibliography{mybib}

\end{document}